\documentclass[preprint,12pt]{elsarticle}

\usepackage{amssymb}
\usepackage{graphicx}
\usepackage{bm}
\newcommand{\be}{\begin{equation}}
\newcommand{\ee}{\end{equation}}
\newcommand{\lra}[1]{\langle #1 \rangle }

\biboptions{authoryear}

\journal{International journal of multiphase flows}

\begin{document}

\begin{frontmatter}



\title{A note on the consistency of Hybrid Eulerian/Lagrangian approach to multiphase flows}


\author[1]{Sergio Chibbaro}
\author[2]{Jean-Pierre Minier}

\address[1]{Institut Jean Le Rond D'Alembert University Paris 6, 4, place jussieu 75252 Paris Cedex 05}
\address[2]{Electricit\'e de France, Div. R\&D, MFEE,  6 Quai Watier, 78400 Chatou, France}

\begin{abstract}
The aim of the present paper is to introduce and to discuss inconsistencies errors that
may arise when Eulerian and Lagrangian models are coupled for the simulations of 
turbulent poly-dispersed two-phase flows. In these hydrid models, two turbulence models
are in fact implicitely used at the same time and it is essential to check that they
are consistent, in spite of their apparent different formulations. This issue appears
in particular in the case of very-small particles, or tracer-limit particles, and it is
shown that coupling inconsistent turbulence models (Eulerian and Lagrangian) can result
in non-physical results, notably for second-order fluid velocity moments. This problem
is illustrated by some computations for fluid particles in a turbulent channel flow
using several coupling strategies. 

\end{abstract}


\end{frontmatter}



\section{Introduction}

Polydispersed  turbulent   two-phase  flows  are   found  in  numerous
environmental and  industrial processes,  very often in  contexts that
involve  additional  issues,   for  example  chemical  and  combustion
ones~\citep{Cli_78}. From
a practical point of view, the Navier-Stokes equations (for the fluid)
and the particle equations (for the particulate phase) must be solved.
Generally  speaking, two basic  approaches have been used for the 
description of the particle phase:  the  {\it Eulerian}, or {\it two-fluid} 
approach, where the  dispersed particle  phase is treated  as a  fluid 
in  much the  same way  as the carrier phase, namely by a set of continuum 
equations which represent the conservation of statistical means, such as 
mass, momentum  and energy, within  some elemental volume of the dispersed 
phase and the so-called {\it Lagrangian}, or {\it particle tracking}, approach 
where individual  particles are tracked through the computed fluid field 
by solving the individual particle equation of motion. In the latter case, 
the complete method for the two phases constitutes a hybrid Eulerian/Lagrangian 
approach. It is worth remembering that various strategies have been explored 
to couple different Lagrangian tracking approaches with Eulerian
approaches for the fluid phase, in particular, with RANS/Moments
approach~\citep{Sto_96,Mur_01,Min_04}, LES~\citep{Boi_00,Jab_02,Bel_04},  and
DNS~\citep{Soldati,Biferale,Toschi,Eaton}.

Let us now introduce the general nature of the Lagrangian approach.  A
no-model approach, in  the spirit  of DNS,  is possible,  but, in
practice, the exact equations of motion are not treatable in realistic
cases.  Indeed, in  the case  of a  large number  of particles  and of
turbulent flows  at high  Reynolds numbers, the  number of  degrees of
freedom turns  out to be  huge and one  has to resort to  a contracted
probabilistic  (modeled)  description.  In  this  case, particles  are
represented by  an ensemble  of Lagrangian stochastic  particles whose
properties are driven by either a model  given in the form of a set of
stochastic  differential  equations (continuous  SDEs)~\citep{Min_01,Chi_08}  or directly  in
terms      of      a       numerical      scheme      (random
walk)~\citep{Sto_96,Pop_87,McI_92}.  The  solution of the
set of stochastic equations represents a Monte Carlo simulation of the
underlying  pdf. Therefore,  this  approach is  equivalent to  solving
directly the corresponding equation  for the pdf in the corresponding
sample-space. In this work, we shall use the  continuous approach. It is 
important to underline  here that this choice is made  for the sake of
clarity  and without  loss  of generality,  since  the Langevin
approach is  expressed in terms  of continuous variables and,  thus, is
physically more intuitive. The random walk models share the same
properties but  they are discrete  and therefore they correspond  to a
numerical scheme for a given continuous stochastic model.

In turbulent two-phase flows, the SDEs equations of the model
contain several mean fields and have the general form
\begin{equation} \label{eq:MK}
dZ_{i}(t) =
A_{i}(t,{\bf Z},\lra{f({\bf Z})})\,dt+
\sum_j B_{ij}(t,{\bf Z},\lra{f({\bf Z})})\,dW_j(t),
\end{equation}   where  the   operator   $\lra{\;}$  stands   for  the
mathematical expectation. 
For exemple, a typical Langevin model has the form~\citep{Min_01}
\begin{eqnarray}
\label{eq:dxp} dx_{p,i} &=& U_{p,i} dt \\
\label{eq:dUp} dU_{p,i} &=& \frac{1}{\tau_p}(U_{s,i} - U_{p,i}) dt \\
\label{eq:dUs}  dU_{s,i}  &=& -\frac{1}{\rho_f}\frac{\partial  \lra{P}
}{\partial x_i}\,  dt +  \left( \lra{U_{p,j}} -  \lra{U_{f,j}} \right)
\frac{\partial   \lra{U_{f,i}}}{\partial   x_j}\,   dt  \nonumber   \\
&-&\frac{1}{T_{L,i}^*}   \left(  U_{s,i}-\lra{U_{f,i}}   \right)\,  dt
  + \sqrt{  \lra{\epsilon}\left( C_0b_i  \tilde{k}/k +
\frac{2}{3}( b_i \tilde{k}/k -1) \right) }\, dW_i~,
\end{eqnarray}
where quantities such as $ T_{L,i}^{*}$, $\tilde{k}$ etc. are defined precisely 
elsewhere~\citep{Min_01,Min_04}. For the sake of present discussion, the important 
point is that this form reveals that different mean fields enter the model equations.
Fluid mean fields, typically $\lra{U_{f,i}}$, are provided by the Eulerian solver 
while particle mean fields, such as $\lra{U_{p,i}}$, are extracted directly from 
the particles and are therefore provided by the Lagrangian solver.
As such, in the case of particles with a non-negligible inertia, the problem is 
well-posed since the different mean fields come from different sources 
(Eulerian/Lagrangian solvers). In that situation, it may even be tempting to believe 
that improving the prediction of one mean field, for example the  fluid mean velocity, 
results in improving the overall capacity of the complete model.

However, the present hybrid method raises issues of consistency between the Eulerian 
and the Lagrangian solvers. This issue is particularly appreciable  in  the  tracer 
limit  (fluid  particle), for which the stochastic two-phase flow model reduces 
simply to a fluid model. This asymptotic limit-case  represents the most relevant 
and fundamental situation where to test the effect of possible errors due to inconsistency, 
for two main reasons: (1) in this limit, as will be shown below,  possible inconsistency can be seen directly at the
level of  fluid moments; (2) one of the main application in bounded flows concerns 
particle deposition, where tracer particle are precisely the most important (aerosols) 
and, often, the least-well predicted by standard models. 
In the particle tracer limit of vanishing inertia, the model (\ref{eq:MK}) takes the form:
\begin{eqnarray}
dx_{f,i} &= &U_{f,i} \,dt\label{eq:modelp} \\ dU_{f,i} &=&
-\frac{1}{\rho}\frac{\partial   \lra{  P   }}   {\partial  x_i}\,   dt
-\frac{1}{T_L}(U_i- \lra{ U_i }^E)dt + \sqrt{C_0\lra{ \epsilon }}dW_i~,
\label{eq:modelv} 
\end{eqnarray}  where $T_L$  represents the  Lagrangian  time-scale of
velocity correlations and it is defined by $  \label{model-TL}   T_L   =  \frac{1}{(1/2   +   3/4
C_0)}\frac{k}{\lra{\epsilon}}~.$ This  model corresponds to  the Simplified Langevin
Model (SLM)~\citep{Pop_94b}.
In the above Langevin model, the fluid mean velocity field appearing in the rhs of the equation, $\lra{ U_i }^E$, is provided by the Eulerian solver as indicated by the index E. 
However, since we are now dealing with fluid particles, the Lagrangian mean velocity field extracted directly from the particles, $\lra{ U_i }^L$, represents the {\it same} physical property. We are therefore in presence of a {\it duplicate field} and the consistency issue requires that $\lra{ U_i }^L= \lra{ U_i }^E$.
Yet, these two mean fields result in fact from two different sources:  
 $\lra{ U_i }^E$ results from the turbulence model chosen in the Eulerian solver whereas 
 $\lra{ U_i }^L$ results from the Langevin model.
Thus, it is  not obvious {\it a priori} to know what happens when
the Eulerian physical  model and the PDF one are not consistent.  For
instance, coupling  Eulerian mean fields computed through  DNS with a
Lagrangian   model   which   is   consistent   with   a   given   RANS
model~\citep{Pop_94b,Mur_99,Mur_01} may introduce a  bias error.  
Unfortunately,
this  point has  not yet received  any  attention  and it has been too quickly believed 
that a "better" mean field  $\lra{ U_i }^E$ fed into the simple Langevin model would automatically
bring about a "better" model.
This route  has been already used  in many works,  notably in particle
deposition  cases~\citep{Kro_00,Mat_00,Tia_07,Deh_08,Par_08,Zha_09}.  Even  though in
some cases and  with some specific models this  procedure may turn out
to be valid, it should be  taken with care in general.  This procedure
is based upon the idea that the Eulerian and the Lagrangian parts of a
hybrid  method  are  completely  independent, in  absence  of  two-way
coupling. It is our purpose to review critically this method.

The question addressed in the present work is:
{\it what is the consequence on the particle simulations of using two inconsistent turbulence models (the turbulence model used in the Eulerian solver and the one corresponding to the Langevin model)? }
In order to develop a quantitative example, the purpose of  this work is to investigate  numerically this effect in a turbulent  channel flow  which is taken here as a
relevant engineering case.

\section{ Model  consistency issue}
Following the discussion in the introduction, it is worth recalling that, 
in  terms of  Eulerian mean
equations,  the  SLM model  is  equivalent  to  the following  turbulence model~\citep{Pop_94,Pop_94b,Min_01}:
\begin{eqnarray}
\label{eq:continuity} &&\frac{\partial  \lra{ U_i }}{\partial  x_i} =0
\\   &&\frac{\partial  \lra{   U_i   }}{\partial  t}   +  \lra{   U_j}
\frac{\partial  \lra{  U_i  }}{\partial  x_j} +  \frac{\partial  \lra{
u_iu_j  }}{\partial  x_j}   =  -\frac{1}{\rho}\frac{\partial  \lra{  P
}}{\partial  x_i} \\  &&\frac{\partial \lra{  u_iu_j }}{\partial  t} +
\lra{   U_k   }   \frac{\partial   \lra{   u_iu_j   }}{\partial   x_k}
+\frac{\partial  \lra{  u_iu_ju_k  }}{\partial  x_k} =  -\lra{  u_iu_k
}\frac{\partial   \lra{  U_j   }}  {\partial   x_k}  -   \lra{  u_ju_k
}\frac{\partial   \lra{   U_i}}  {\partial   x_k}   \nonumber  \\   &&
\hspace{7cm}  -\frac{2}{T_L}\lra{  u_iu_j  }  +  C_0\lra{  \epsilon  }
\delta_{ij}.
\label{eq:rans}
\end{eqnarray}   Using   the   expression   retained  for   $T_L$   in
Eq.~(\ref{eq:modelv}),  the  transport  equation for  the  second-order
moments can be re-expressed as :
\begin{eqnarray} \frac{\partial \lra{  u_iu_j }}{\partial t} + \lra{
U_k  } \frac{\partial  \lra{ u_iu_j  }}{\partial  x_k} &+&\frac{\partial
\lra{ u_iu_ju_k }}{\partial x_k} = -\lra{ u_iu_k }\frac{\partial \lra{
U_j  }} {\partial  x_k}  - \lra{  u_ju_k  }\frac{\partial \lra{  U_i}}
{\partial    x_k}    \nonumber    \\    &-&  (1    +
\frac{3}{2}C_0)\left( \lra{u_iu_j}  - \frac{2}{3}k \delta_{ij} \right)
- \frac{2}{3}\delta_{ij}\lra{ \epsilon }.
\end{eqnarray}   This   shows   that   the  SLM   corresponds   to   a
$R_{ij}-\epsilon$ Rotta model~\citep{Pop_94b}. We have
retained this simple  version, namely the SLM model,  which is consistent
with usual  Reynolds-stress models  as a kind  of sound basis  for the
numerical  investigations on  particle deposition  though it  is clear
that,  at least  for the  prediction  of fluid  mean quantities,  this
leaves room for improvement by using more complex Langevin ideas.

\section{Numerical                                             Results}

A stationary  turbulent channel-flow is solved first  with the Eulerian
method  and, then,  with  the  PDF one.   The  computations have  been
performed for  the case of  fully developed turbulent channel  flow at
$Re =  u_{\tau}h/\nu =  395$.  The channel  flow DNS results  of Moser
{\it et al.} \citep{Mos_03} have been taken for comparison as reference
data.  For  the simulations,  the mesh used  in DNS (128  points along
$y$) is also used here, which assures a good interpolation of the mean
fields for  the Lagrangian stochastic  particles.  Quantities designed
with   the  upper-script   $+$  are   non-dimensionalised   with  wall
parameters.  Being  a particle-mesh method, particles move in a
mesh where,  at every cell, the  mean fields describing  the fluid are
known.   Generally   speaking,  the  statistics   extracted  from  the
variables attached to  the particles, which are needed  to compute the
coefficients of system (\ref{eq:modelp})-(\ref{eq:modelv}) are not
calculated for each  particle (this would cost too  much CPU time) but
are evaluated at  each cell center following a  given numerical scheme
(averaging operator). The stochastic equations  (\ref{eq:modelp})-(\ref{eq:modelv}) are
solved   through  a   consistent  first-order,   unconditionnally  stable
numerical scheme~\citep{Min_03}.  No-slip and impermeability conditions are satisfied
by imposing boundary conditions on the stochastic particles.  Symmetry
conditions are  imposed at the other  boundary placed a  the center of
the channel $y=h/2$.  Along the $x$ direction, periodic conditions are
imposed.  The details of the numerical approach have been exhaustively
described in a recent article~\citep{Pei_06}.

\subsection{Consistency}\label{sec:cons}
\begin{figure}[h]
\begin{center}
\includegraphics[scale=0.3]{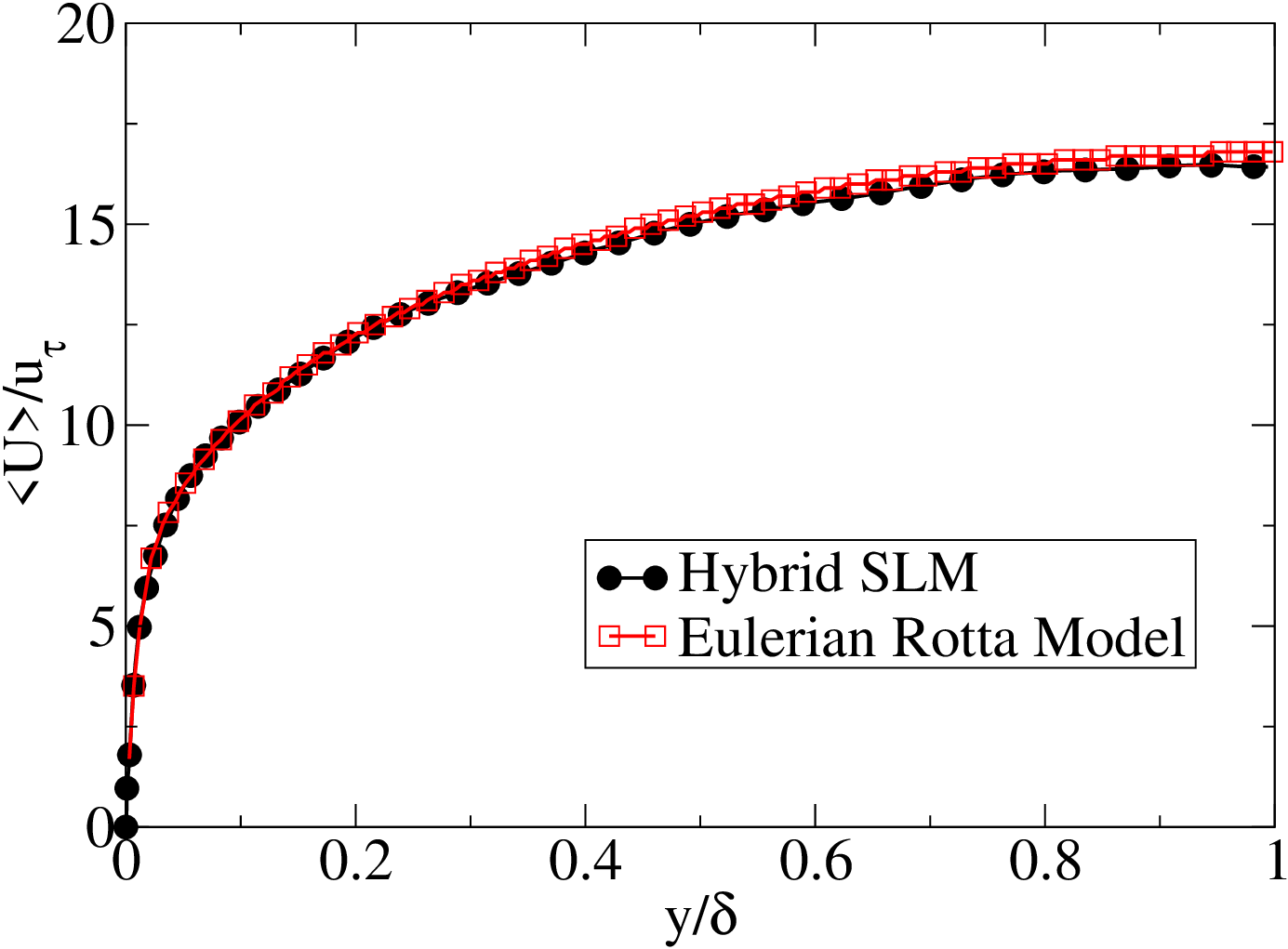}\vspace{1cm}
\includegraphics[scale=0.3]{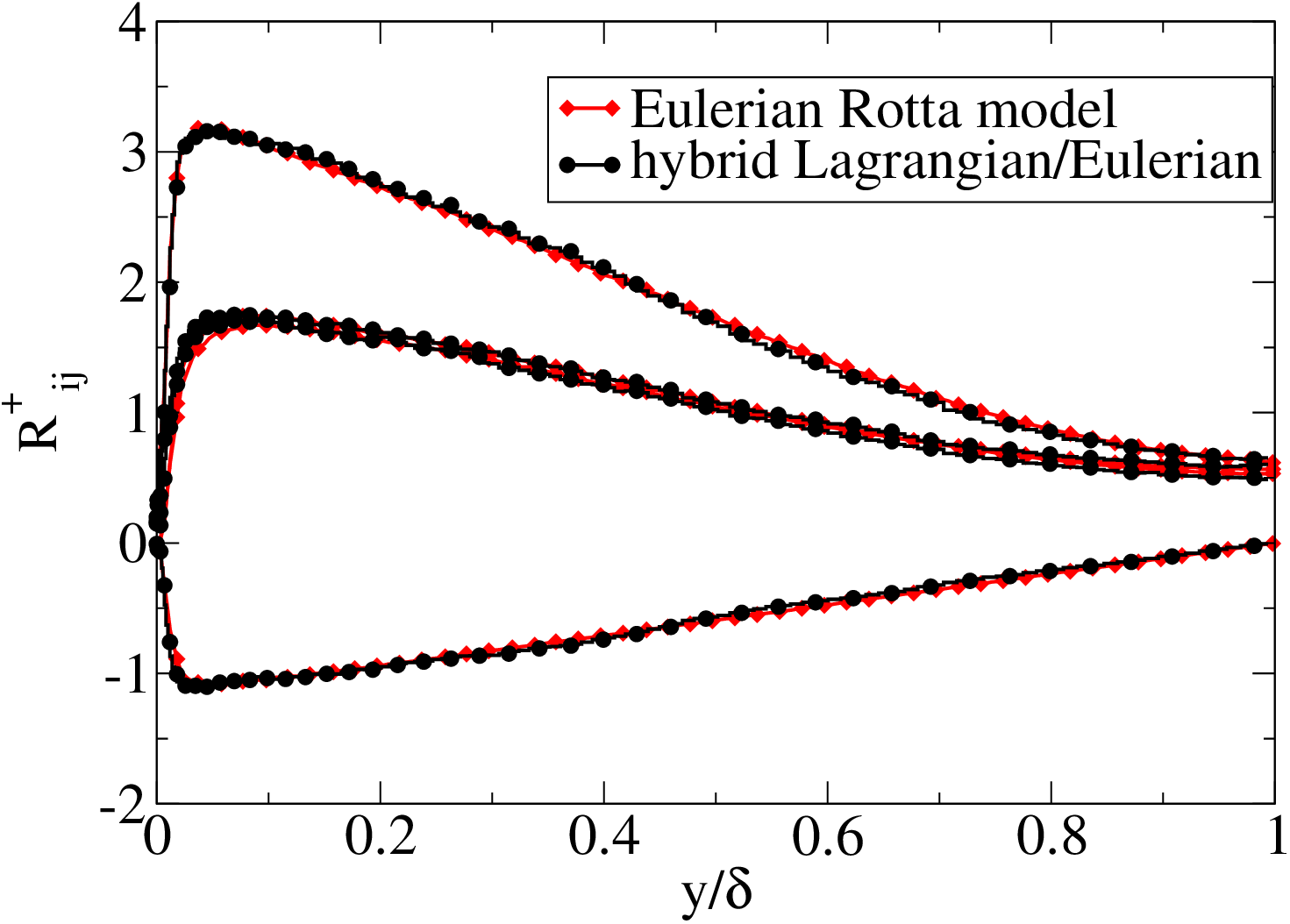}
\caption{Mean velocity (a) and Reynolds stress (b) results obtained in
the Eulerian method, in the present Hybrid configuration
and  in the  DNS~\citep{Mos_03}.   The results  are in  non-dimensional
units.}  \vspace{1cm}
\label{Fig_1}
\end{center}
\end{figure}  

The  first  set   of  numerical  experiments  has been
performed coupling  the Eulerian  Rotta Reynolds-stress model  with the
Langevin  Simple   model  previously  described. Identical results 
are expected, since the  equations for the moments  of first and  the second order
are the same for  both models and, thus, consistency
is  assured.  In figures \ref{Fig_1}a, the mean-velocity  profiles obtained with 
the present hybrid configuration are shown.  In figures
\ref{Fig_1}b,  the  profiles  of  second-order moments,  that  is  the
Reynolds stress  tensor, are also shown,  for the present hybrid and for the
Eulerian calculations.  The Eulerian and Lagrangian profiles are in a quite
good agreement.  In conclusion, the mean fluid velocity moments derived from the present
PDF  method  are in  agreement  with  those computed  in  the
Eulerian  configuration and  thus with  the physical  expected values.
Moreover, the  global hybrid method  Eulerian (Rotta-model)/Lagrangian
(SLM)  is  demonstrated  to  be  consistent.   It  is  worth
emphasising  that the ingredients  that have  been necessary  to reach
this objective are:
\begin{enumerate}
\item[(i) ] Consistent physical model.
\item[(ii)] Consistent numerical scheme
\item[(iii)] Accurate  global numerical  method , concerning  also the
exchange of information from Eulerian solver to Lagrangian one.
\end{enumerate}

\subsection{Hybrid Consistency error}

We can  now study the global  error which is  possibly introduced by
using a hybrid method not completely consistent.  Generally speaking,
hybrid Eulerian/Lagrangian  methods are affected by  different kind of
errors due  to: (1)  spatial discretisation; (2) time discretisation;
(3) the use  of a finite number of particle  and per cell (statistical
and bias errors).   All these errors have been  made negligible in the
following simulations, in order  to isolate the ``hybrid consistency''
error.  In both  methods, a time-step of $10^{-4}s$  and a spatial-step
of $10^{-4} \delta$ have  been used, which have been shown in numerical simulations to be sufficiently small for our
purposes.   $5*10^{4}$  particles have  been  employed,  which can  be
considered high enough in the present case.  For the sake of clarity, 
the configuration discussed
in  the last section,  where Eulerian  mean variables  consistent with
Lagrangian ones were used, will be called the {\it standard} configuration
in the following.  \vspace{0.75cm}
\begin{figure}[ht]
\begin{center}
\includegraphics[scale=0.33]{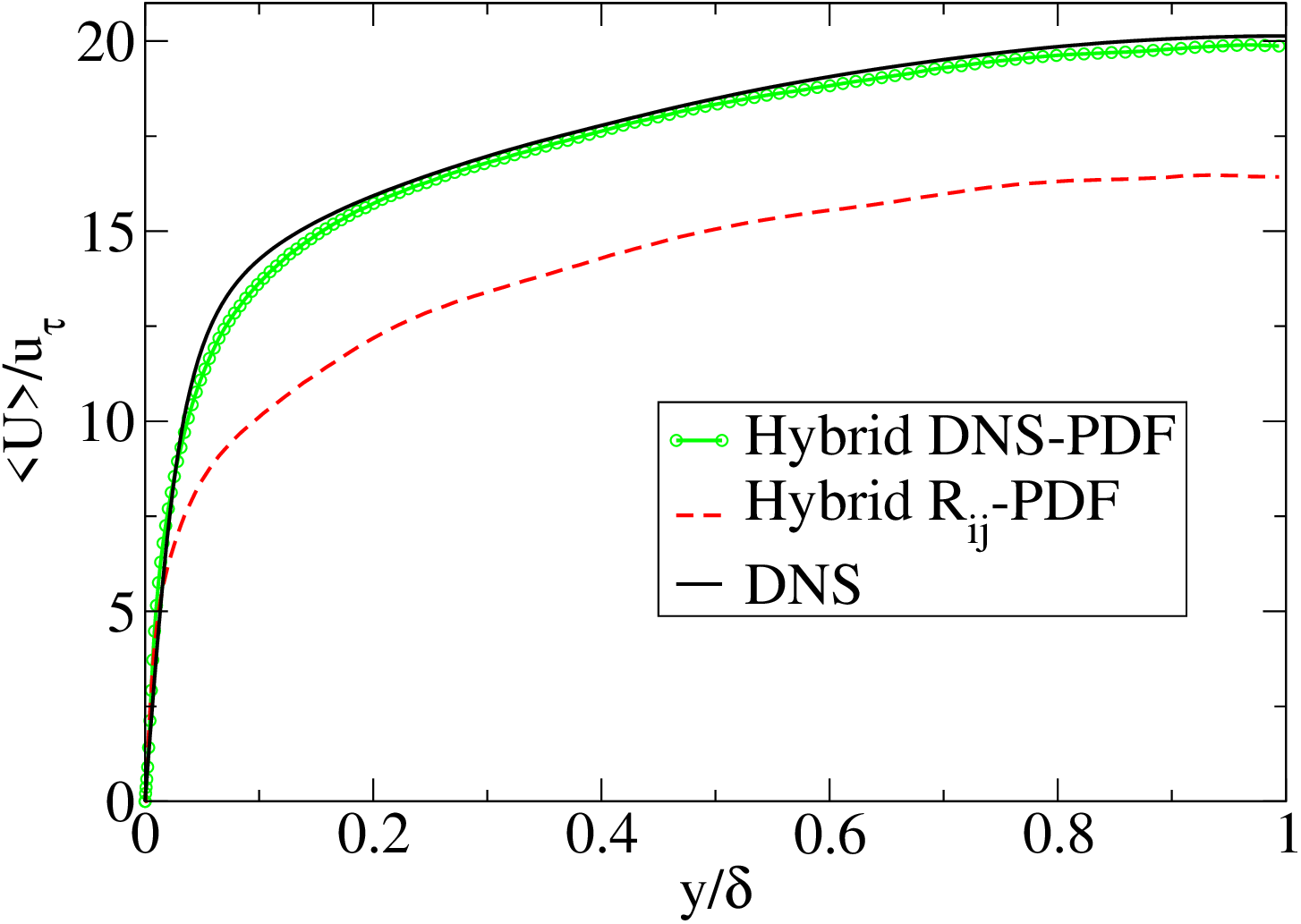}
\includegraphics[scale=0.33]{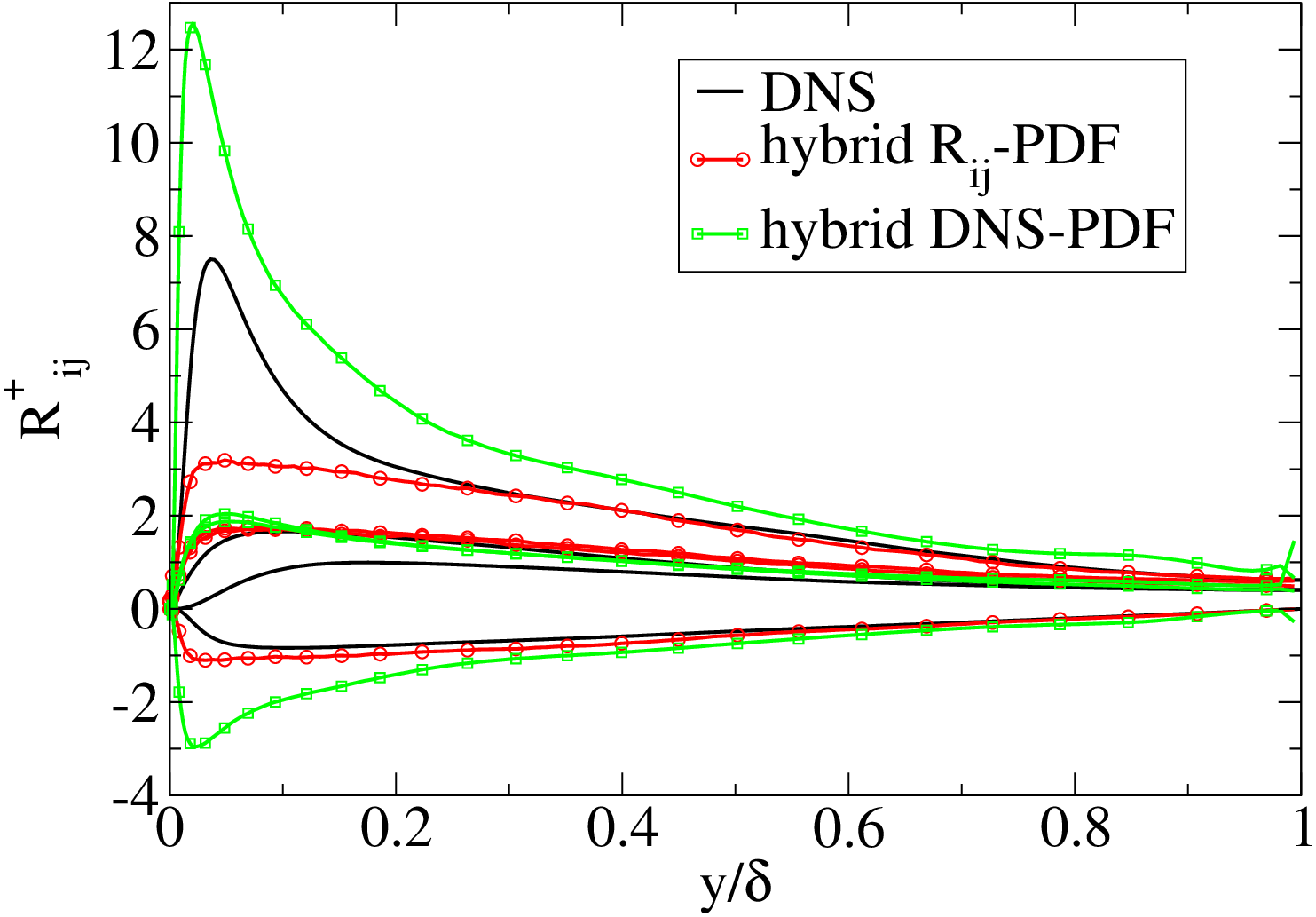}
\caption{DNS-PDF  configuration:  (a)  mean  velocity.   (b)  Reynolds
stress. DNS~\citep{Mos_03} and $R_{ij}$/PDF  profiles are also give for
comparison.}
\label{Fig_3}
\end{center}
\end{figure}  

First,  let  us  consider the  following  configuration: an Eulerian 
DNS~\citep{Mos_03} is now coupled to the present  Lagrangian method (this means 
that the mean fluid velocity $\lra{ U }^E$ provided to the SLM model is
given by a DNS calculation). In figure~\ref{Fig_3}a, the mean velocity computed 
by the PDF method in this
configuration, together with DNS original  profiles as well as the mean
velocity  computed  by  PDF  method  in  the  consistent  $R_{ij}$-PDF
configuration, are  shown.  The mean  velocity obtained in  the DNS/PDF
configuration is in good agreement with the DNS one and, thus, is strongly
different from the one obtained  in the {\it standard} configuration.  In
fact, the exact profile is now recovered.  In figure \ref{Fig_3}b, the
Reynolds stress  are shown for the same  configurations.  The profiles
are now dramatically  changed in  comparison with the {\it standard} results.
The $\lra{uv}$ and the $\lra{u^2}$ components are strongly overpredicted.  
On the contrary, the other
diagonal  components   do  not  present   appreciable  changes.   This
behaviour can  be explained.  As diagonal  $\lra{v^2}$ and $\lra{w^2}$
components are  essentially dependent on Lagrangian  time-scale and on
diffusion coefficient, they are independent of
the  Eulerian model  and  therefore not  affected  by the  consistency
error.  On the contrary,  the cross-shear stress $\lra{uv}$ depends not only
on the turbulent kinetic energy and the Lagrangian time-scale but also explicitely
on the gradient  of the  mean velocity. It is therefore much more sensitive
to the prediction of the Eulerian mean velocity. The present mean velocity profile
attains a higher maximum than in the {\it standard} configuration and,
thus,  it is  much steeper.   The shape  of $\lra{u^2}$  is in  turn a
direct consequence of this behavior.   It is important to stress here
that DNS/PDF  Reynolds stress profiles are  unphysical; notably, while
the negative  peak of  the cross-shear stress  {\em should} be  of the
order of 1, it turns out to be $\lra{uv}\approx-3$. \vspace{0.75cm}
\begin{figure}[h]
\begin{center}
\includegraphics[height=6cm,width=8cm]{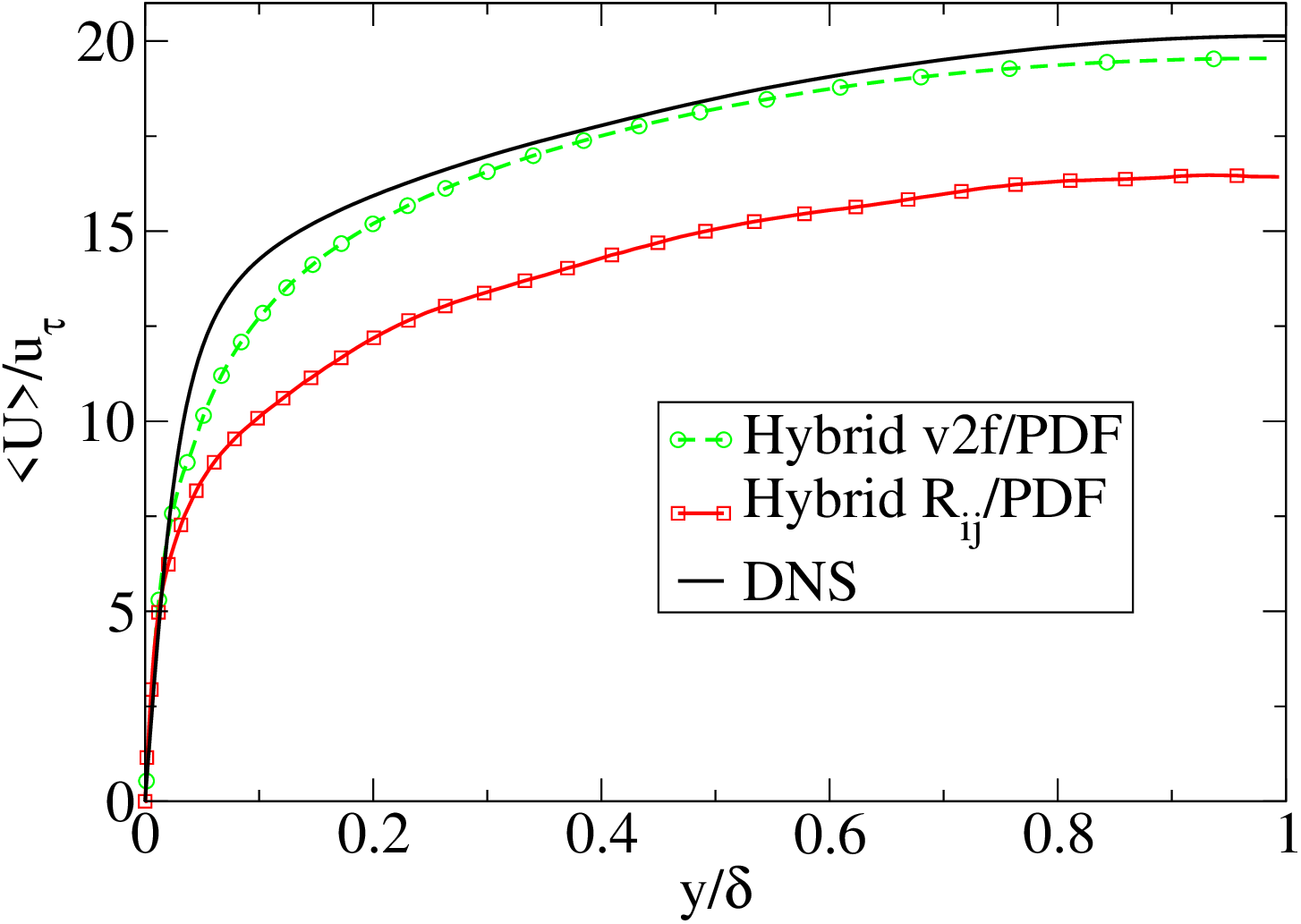}
\includegraphics[height=6cm,width=8.5cm]{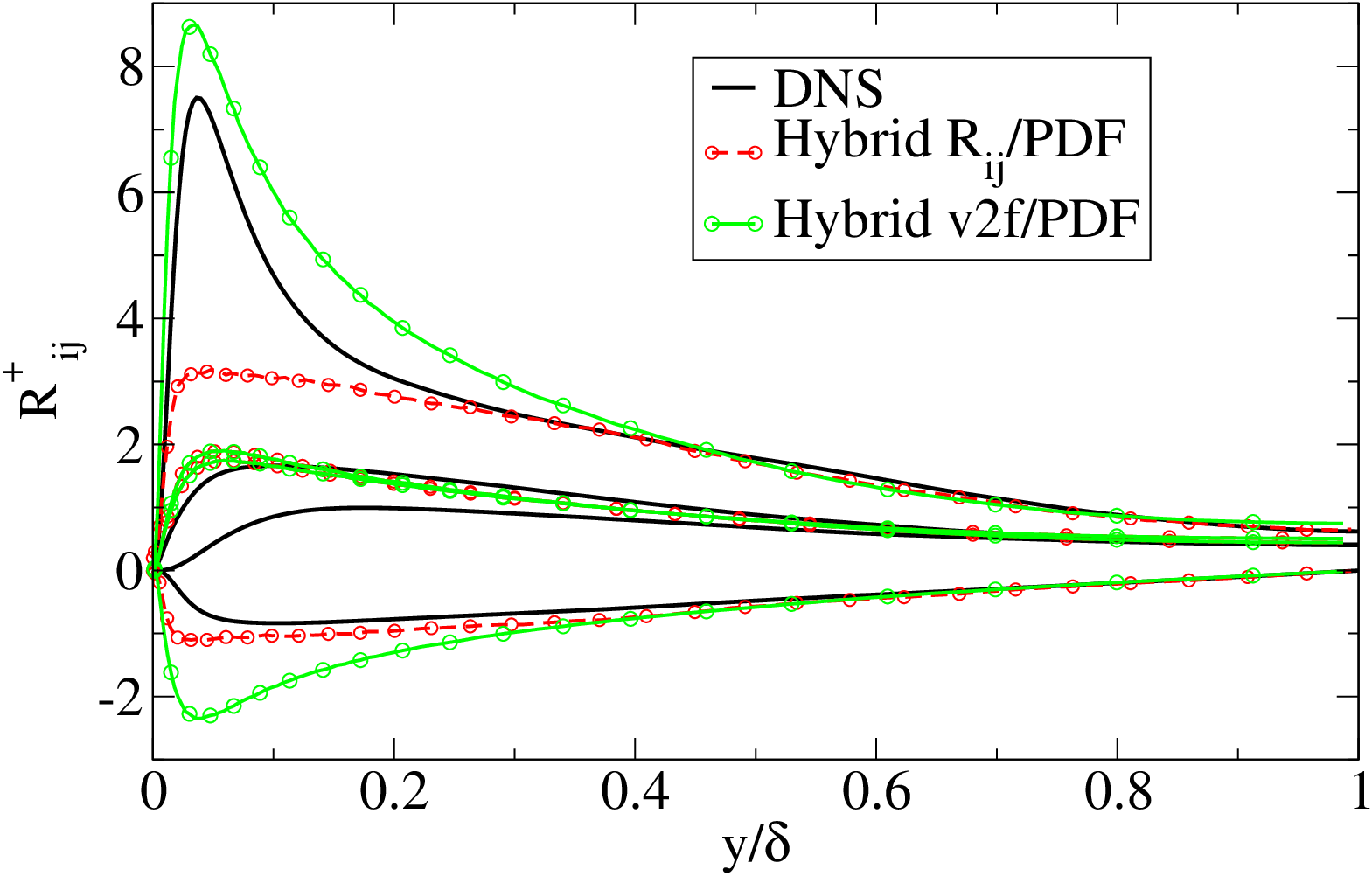}
\caption{v2f-PDF  configuration:  (a)  mean  velocity.   (b)  Reynolds
stress. DNS~\citep{Mos_03} and $R_{ij}$/PDF  profiles are also give for
comparison.}
\label{Fig_4}
\end{center}
\end{figure} 

Finally, we  analyse the  results obtained  by coupling the
present PDF  model to a low-Reynolds  number RANS model  known for his
good performance  in boundary flows, the v2f  model \citep{Dur_00}.  In
the present  calculations,  we  use  a refined version  of  the  model
\citep{Lau_05}.   In figure  \ref{Fig_4}a,  we show  the mean  velocity
computed  in this  hybrid  configuration together  with  the DNS and  {\it
standard} PDF results.  As previously noted,  the mean velocity given by the PDF
method is approximately equal to the Eulerian mean one provided to the PDF
solver, in this case computed by v2f solver.  Furthermore, this result
is also  in good agreement  with DNS result.  In  figure \ref{Fig_4}b,
the Reynolds  stress  profiles  for  the same  configurations  are  shown.
The results  are   similar  to  those  obtained  in   the  hybrid  DNS/PDF
configuration.   Even  though   the  $\lra{uv}$  and  the  $\lra{u^2}$
components are  less overpredicted,  they still show qualitatively  the same
behaviour and  remain unphysical.   The consistency error  is slightly
less important but still large.

Some observations are in order:
\begin{itemize}
\item[(i)] In all configurations, the mean velocity computed from stochastic
particles basically collapses on the  value given by the Eulerian mean
velocity  used in  the Lagrangian  model.  This  is in  line  with the
physics   of    the   Langevin   model,   which   is    based   on   a
return-to-equilibrium idea \citep{Min_97}.
\item[(ii)]  In  the PDF-DNS  and  PDF-V2F  configurations,  a  large
difference  between the Eulerian and  the Lagrangian results  at the  level of
the second-order statistical moments (Reynolds  stress) is found.  This is
a  direct consequence  of coupling, in a hybrid approach, two methods
which  are   not  consistent.   Thus,  the hybrid   consistency  errors  are
identified by comparing the actual Reynolds stress profiles with those
computed    in   the  {\it standard}  configuration,    see
figs. \ref{Fig_3}b,\ref{Fig_4}b.
\item[(iii)] Results obtained in the hybrid DNS/PDF and v2f/PDF approaches
are  qualitatively similar,  even though the DNS and the v2f  approaches are
quite  different from  a theoretical  point of  view.  This  should be
expected.   The  profiles  provided  by  the Eulerian  solver  to  the
Lagrangian one are the mean velocity, the mean pressure, the turbulent energy and
the turbulent dissipation.   For these  variables, the v2f  approach gives
results which are in good agreement with the DNS ones. Therefore, from the point of
view of  the hybrid  method DNS and  v2f approaches  are qualitatively
similar.   Moreover, from a  quantitative point  of view,  the results
obtained through  the hybrid v2f/PDF approach are  in better agreement
with  those calculated  in  the {\it standard}  $R_{ij}$/PDF case  than
those  obtained  using   the  DNS/PDF  approach,  fig.   \ref{Fig_4}b.
Therefore,  v2f model  is found  to  be more  consistent with  present
Lagrangian model than DNS and the hybrid consistency error is smaller.
\end{itemize}

\section{Conclusions}

A  study of  a turbulent  channel  flow has  been carried  out in  the
framework  of  a  Hybrid  Eulerian/Lagrangian approach with the main attention
pointed to  the issue of  the consistency between the  two approaches.

We have  used different  kind of Eulerian  methods (RANS  $R_{ij}$ and
v2f,  DNS)  coupled  with  the  simplified Langevin  model  for  the
Lagrangian tracking of fluid particles.  It is worth underlying that
this kind of couplings are normally used 
in multiphase simulations, for instance DNS~\citep{Mat_00,Tia_07,Deh_08},
v2f~\citep{Zha_09} among others.
The Lagrangian model proposed
is  in  the  form  of  a  set  of  stochastic  differential  equations
(\ref{eq:modelp})-(\ref{eq:modelv})  and, in  practice,  the PDF  underlying
this stochastic process is obtained  via a Monte Carlo method, that is
through  the simulation  of a  number of  stochastic  particles.  This
feature is  very useful to recognize immediately  the physical contain
of  the model, at  variance with  other more  heuristic but  also very
popular  approaches like  the  discrete random  walk  models.  

We  have analysed the changes produced in the Lagrangian results by 
choosing different  Eulerian methods.  In particular,  we have studied
the duplicated  variables which are the  two first moments  and we have
shown that the  global method is found to  be consistent when $R_{ij}$
Rotta model is coupled with the SL model, as it is theoretically expected.
On the contrary, results obtained in other configurations give evidence of
a bias error  precisely due to the inconsistency  between the Eulerian
and Lagrangian physical  models.  It is worth emphasizing  here that it
is possible to assess the global consistency of the method, because we
use a  completely consistent and  an accurate numerical  scheme, which
allows us to consider the numerical errors as negligible.

These results underline  that, in hybrid Eulerian/Lagrangian approaches
to turbulent  flows, delicate questions of consistency  between the two
parts of the method ineluctably arise and deserve a careful treatment.
In  general, the  simple statement  that a "better fluid
profile"  help  to improve Lagrangian  results (independently
of the Lagrangian  model) is not true.  At variance with this belief, using an
Eulerian  model  which  is  not  consistent with  the  Lagrangian  one
introduces  a  consistency error  which can introduce a flaw in the global  method.

Finally, in  authors' opinion, the  Lagrangian tracking method  is too
often considered  as a {\em  black box} tool  scarcely important  from a
physical  point  of view  and  thus  hierarchically  subjected to  the
Eulerian method.   In fact, we  think that in  hybrid Eulerian/Lagrangian
approaches to two-phase flows  the Lagrangian part is more fundamental
and has  the guiding  role. In  this sense, our  results show  that it
could be  wise to start from  the choice of the  Lagrangian model and
afterwards to choose an Eulerian model which is consistent with it.

\bibliographystyle{elsarticle-harv}



\end{document}